# Magnetic Order in the Pyrochlore Iridates $A_2Ir_2O_7$ ($A$ = Y, Yb)

S. M. Disseler[1], Chetan Dhital[1], A. Amato[2], S. R. Giblin[3], Clarina de la Cruz[4], Stephen D. Wilson[1], and M. J. Graf [1]*

[1] Department of Physics, Boston College, Chestnut Hill, MA 02467, USA

[2] Paul Scherrer Institute, CH 5232 Villigen PSI, Switzerland

[3] Rutherford Appleton Laboratory, Didcot, Oxfordshire OX11 0QX, UK

[4] Quantum Condensed Matter Division, Oak Ridge National Laboratory, Oak Ridge, TN 37831-6393, USA

**Abstract**

We present results from muon spin relaxation/rotation, magnetization, neutron scattering and transport measurements on polycrystalline samples of the pyrochlore iridates $Y_2Ir_2O_7$ (Y-227) and $Yb_2Ir_2O_7$ (Yb-227). Well-defined spontaneous oscillations of the muon asymmetry are observed together with hysteretic behavior in magnetization below 130 K in Yb-227, indicative of commensurate long-range magnetic order. Similar oscillations are observed in Y-227 below 150 K; however the onset of hysteretic magnetization at $T$ = 190 K indicates a transition to an intermediate state lacking long-range order as observed in Nd-227. Our results also show that insulating members of the iridate family have nearly identical magnetic ground states, and that the presence of magnetic $A$-site species does not play any significant role in altering the ground state properties.

## I. Introduction

Transition-metal oxides of 5d elements have been the growing focus of a large number of studies, primarily due to the interplay between electron-electron correlations and spin-orbit interactions in these materials. Iridate compounds forming in the pyrochlore lattice have been of particular interest as they are predicted to host novel topological phases such as Dirac or Weyl semi-metal and axion insulator phases [1-4], which are tuned by the strength of these correlations. The Dirac semi-metal offers a new possibility of realizing topologically protected conduction states at points in the bulk of 3D materials in contrast to surface topologically insulating (TI) behavior observed thus far [1,2]. The pyrochlore family of iridate compounds, $A_2Ir_2O_7$ ($A$ = Y, lanthanide), or $A$-227, is particularly promising to observe these and other predicted behaviors as the correlation strength can be varied substantially by choice of $A$-site ion, from magnetically ordered-insulator [5,6] to complex metallic ground state [7-9].

The magnetic structure of these materials is of considerable interest as it relates directly to the topology of the ground state [1,2]. Trigonal crystal fields at both the $A$ and Ir sites create Ising-like moments that can order in a non-collinear arrangement with moments along the local (111) directions. Earlier works on insulating $A$-227 revealed bifurcation of the magnetic susceptibility above 100 K near a metal-insulator transition, $T_M$, which was interpreted as the formation of antiferromagnetism [6, 9, 10]. The appearance of spontaneous oscillations in muon spin relaxation/rotation (μSR) measurements of Eu-227 below this temperature provided convincing evidence of a transition to a state with long-range order (LRO) [5]. As $Eu^{3+}$ has a $J = 0$ Hund's rule ground state, the oscillations can be attributed to magnetic order of the $Ir^{4+}$ moments. Similar measurements on the weakly-metallic Nd-227 compound show strongly damped oscillations below $T = 8$ K, indicative of an ordered ground state with a large internal field distribution [8], while Pr-227 did not show any such oscillations to 30 mK [11]. Despite this, elastic neutron scattering has yet to provide conclusive evidence for LRO on the Ir-sites in any of these systems [8,12,13]. It is thus important to characterize additional members of the $A$-227 family to gain insight into the competing interactions in this system, the resultant magnetic ground states, and the prospects for observing novel topological states of matter.

This work presents results from resistivity, magnetization, neutron scattering, and μSR experiments on polycrystalline samples of two members of the iridate family, $Y_2Ir_2O_7$ (Y-227) and $Yb_2Ir_2O_7$ (Yb-227). We find long-range order in both compounds, akin to that observed in Eu-227, as evident from the observation of zero-field muon spin precession, with an apparent mean field behavior below the onset temperatures of $T$ = 150 K and 130 K for Y-227 and Yb-227, respectively. However, the onset of hysteretic magnetization at 190 K for Y-227 does not coincide with the onset of oscillations in μSR. This is indicative of a transition to a glassy or frustrated intermediate phase as observed in Nd-227 [8], and points to the critical role of the correlation energy in stabilizing this state. As Y-227, Yb-227, and Eu-227 are all insulating at low temperatures, our results indicate these materials exhibit similar ground state arrangements of $Ir^{4+}$ moments, independent of the correlation energy or magnetic properties of the $A$-site element.

## II. Materials and Experiment

Polycrystalline samples of $A_2Ir_2O_7$ ($A$ = Y, Yb) were synthesized by reacting stoichiometric amounts of $A_2O_3$ (99.99%) and $IrO_2$ (99.9%). Powders were pelletized using an isostatic cold press and reacted at temperatures between 900 C to 1125 C for a period of six days with several intermediate grindings. The samples were determined to be nearly phase-pure with the exception of two minor impurity phases of $IrO_2$ and $A_2O_3$ ($A$ = Y, Yb) each individually comprising less than ~1% of the total volume fraction. X-ray diffraction measurements were performed on a Bruker D2 Phaser diffractometer on both of the above two samples at room temperature for crystal structure determination. Using the cubic Fd-3m space group, the lattice parameters were found to be $a$ = 10.1699(4) Å for $Y_2Ir_2O_7$ and 10.1015(5) Å for $Yb_2Ir_2O_7$. Polycrystalline $Y_2Ir_2O_7$ was also probed via neutron diffraction measurements inside of a closed cycle refrigerator at the High Flux Isotope Reactor (HFIR) at Oak Ridge National Lab on the HB2-A diffractometer. Data from HB-2A were collected with $\lambda_i$=1.5385 Å, a Ge(115) monochromator, and 12’-31’-6’ collimations. All diffraction data were refined using the FullProf software package. The XRD and neutron diffraction patterns along with refinement can be seen in Figure 1.

Electrical resistance was measured using a standard four-terminal ac bridge technique. Bulk magnetization was measured between 300 K and 2 K using a Quantum Design MPMS. The magnetization data were taken during warming in a 1 kOe applied field preceded by cooling from 300 K to 2 K in either zero field (ZFC) or 1 kOe (FC).

Muon spin relaxation (μSR) experiments over the range $1.5 < T < 220$ K in zero applied magnetic field (ZF) were carried out on the pulsed $\mu^+$ source at ISIS using the EMU spectrometer, and on the πM3 beam line at the continuous source at PSI using the GPS spectrometer. In addition, limited measurements were performed in longitudinal fields (LF) up to 3 kG at ISIS. The two spectrometers are used in a complimentary fashion to measure both fast (GPS) and slow (EMU) muon depolarization processes. Samples at ISIS were mounted on a silver-backing plate that contributed a small amount to the total asymmetry (<10%), which was independently measured and subtracted. The same powders were sealed inside thin metalized Mylar foil for use in GPS, eliminating any such background contributions.

## III. Results

The temperature dependent static susceptibilities, $\chi = M/H$, for Y-227 and Yb-227 are shown in Fig 2(a) and (b) respectively. Under ZFC conditions we observe a peak for Y-227 at T = 180 K, while FC conditions show a sudden increase beginning at 190 K, and a second inflection near 150 K. A similar bifurcation is observed in $\chi$ for Yb-227 at 135 K, as seen in the inset of Figure 2(b), although it is superimposed on a much larger paramagnetic component. The difference between ZFC and FC below the bifurcation temperature is of similar magnitude in both samples (of order $10^{-2}$ emu/mol), suggesting that the bifurcation arises from the Ir-sublattice, with an additional large paramagnetic contribution in Yb-227 due to the weakly coupled $Yb^{3+}$ ($J = 7/2$) moments.

The temperature dependence of the zero field resistivity is shown in Figures 2 (c) and 2(d) for Y-227 and Yb-227, respectively. Both samples exhibit monotonically increasing resistivity with decreasing temperature, with resistivities of approximately 30 Ω-cm at room temperature, and which exceed 1 MΩ-cm by 10 K. The resistivity does not follow a thermally activated exponential behavior as expected for simple semiconductors, rather an unusual power law behavior appears below 150 K for Y-227 and 130 K for Yb-227 with $\rho \sim T^{-4}$ for both compounds (as shown on the double log-plot in the inset of 2(c) and 2(d)). We note that several samples from different batches yielded identical results; additionally, a sample of Y-227 was annealed in a flowing $O_2$ at 1000 C, again with no change in the observed behavior. These samples exhibit much stronger insulating properties than previously reported [10], most likely due to reduced impurity carrier contributions and/or a more uniform oxygen stoichiometry.

Elastic neutron scattering measurements were carried out as an initial probe of $Ir^{4+}$ order for Y-227. Structural refinement of the data (Fig. 1(c)) showed excellent agreement with previously reported values; we did not observe any evidence of long-range order or structural transitions down to $T = 3$ K. However neutron studies of small moment iridates often require single crystal measurements to resolve correlated spin scattering. Conservatively, this powder measurement places an upper limit of the $Ir^{4+}$ ordered moment to be below 0.5 $\mu_B$ based on the collected statistics and the resolution of the diffractometer.

μSR provides an excellent means of further probing the local magnetic order due to large gyromagnetic ratio of the muon ($\gamma_\mu/2\pi$ = 135.5 MHz/T) which makes it possible to detect static fields of few Gauss or less while simultaneously probing spin dynamics and correlations in the MHz regime [14]. Example asymmetry curves showing the early time behavior from PSI are shown in Fig. 3(a) for Y-227 and Fig. 4(a) for Yb-227. The most significant feature is the appearance of spontaneous muon spin precessions below $T$ = 150 K and 130 K for Y-227 and Yb-227, respectively. These data signify the presence of a static local magnetic field $\langle \boldsymbol{B_{loc}} \rangle$ at the muon site. Furthermore, the oscillations are well defined indicating commensurate order with a single magnetically unique muon stopping site. The resultant asymmetry curves were fit by the simple depolarization function for a magnetically ordered polycrystal, as used in Ref [5] for Eu-227:

$$A(t) = A_1 \exp\left[-(\Lambda t)^{\beta}\right] \cos(\omega t + \phi) + A_2 \exp(-\lambda t) \quad . \quad (1)$$

The first component describes the oscillations from muon spin precessing about a spontaneous static local field with frequency $\omega_\mu/2\pi = \langle \boldsymbol{B}_{loc} \rangle \gamma_\mu/2\pi$ and damping described by a stretched exponential with characteristic rate $\Lambda$ and $\beta < 1$. The second component describes the relatively slower longitudinal relaxation due to spin-lattice relaxation or fluctuations of the local moments. Because this component also reflects muons for which the initial muon polarization is parallel to the internal field at the stopping site, we expect $\eta = A_2/(A_1 + A_2) = 1/3$ at temperatures well below the ordering temperature, and indeed we find that $\eta = 0.30(2)$ and 0.35(1) for Y-227 and Yb-227, respectively, when $T = 1.8$ K. The phase factor $\phi$ was found to be rather large for a system described by a single oscillatory frequency; however suitable fits for both samples at all temperatures were obtained by fixing $\phi$ to -20°, the value obtained at 1.5 K.

For Y-227, the temperature dependence of $\omega_\mu$ as determined from fitting to Eq. 1 is shown in the top panel of Fig. 5. The onset of muon spin precession occurs near 150 K, and the precession frequency increases monotonically, reaching a maximum value of $\omega_\mu/2\pi = 14.5(1)$ MHz at 1.6 K, corresponding to a low temperature value of $\langle \boldsymbol{B}_{loc} \rangle =$ 1100(10) G. This variation is mirrored in the temperature dependence of the peak frequency extracted from Fast Fourier Transforms (FFT), as shown in Fig 3(b). The resultant temperature dependence of the extracted frequency below 150 K is representative of the magnetic order parameter and is described by the phenomenological mean-field expression $\langle \boldsymbol{B}_{loc} \rangle \sim (1-(T/T_C)^{\alpha})^{\beta}$. Shown as the solid line in Figure 5, fitting to this expression yields values of $\alpha = 4.1(1)$ and $\beta = 0.29(2)$. The value of $\beta$ is close to that expected for 3D systems [15], however the value of $\alpha$ is much larger than the value ~2 usually found for spin-wave excitations [16], although this difference may result from an insufficient density of points in the vicinity of the transition.

The appearance of the spontaneous precession occurs at a temperature significantly below the bifurcation temperature. We believe this behavior is intrinsic to the sample and indicative of a transition from a intermediate temperature magnetic phase with short-range order to a phase with long range order. XRD and neutron scattering have ruled out the presence of impurity phases above 1% volume fraction (and both impurity phases do not order magnetically), and μSR is a volumetric probe, so the success of Eq. 1 in describing the low-temperature depolarization curves, along with the longitudinal field results described below, rule out any coexistence of two phases at low temperatures.

Dephasing of the spontaneous muon spin precession is described by the phenomenological damping parameter $\Lambda$, shown in the middle panel of Fig. 5 for Y-227. The sharp increase of $\Lambda$ just below $T_M$ is likely an artifact of the rapidly vanishing contribution to the asymmetry of the oscillatory component rather than critical divergence, as no such increase is seen in the paramagnetic region above $T_M$. The fractional quantity $\Lambda/\omega$ was found to be ~ 0.1 and constant at low temperatures, verifying that the internal field is fairly uniform at each muon stopping site except at temperatures near $T_M$. The stretching exponent $\beta$ was found to be 0.6(1) and roughly temperature independent for Y-227 for temperatures below 120 K.

Shown in Fig 6, the onset of spontaneous muon spin precession is also observed in Yb-227 below $T$ = 130 K, with a monotonic temperature dependence down temperatures near 20 K, with an extracted frequency $\omega_\mu/2\pi$ = 14.4(1) MHz ($<\boldsymbol{B_{loc}}>$ = 1100 G) at 20 K. Below a characteristic temperature T* ~ 20 K, however, the frequency is observed to increase again from 14.4 MHz to 15.5 MHz at 1.6 K, indicating the onset of a second contribution to the local field. Correspondingly, a second increase of $\Lambda$ is observed below $T^*$ indicating an increased width of $<\boldsymbol{B_{loc}}>$.

The FFTs of the asymmetry curves, seen in Fig 4(b), show that below $T^*$ there is an increase in both the location and width of the peak and confirm the onset of an additional static local field. We do not observe any features in the magnetization near this temperature that would suggest a change in the magnetic order, and no such feature is observed in either Y-227 or Eu-227, which have non-magnetic ions on the *A*-site sublattice. A similar phenomenon was observed in the series *RFeAsO*, in which the localized rare earth moments become polarized via hyperfine or exchange interactions with the ordered Fe lattice [17]. We find the temperature dependence of the muon spin precession frequency for Yb-227 is well described by the same modified mean-field form used in Ref 17 to describe CeFeAsO:

$$f = f_0\left[1-\left(\frac{T}{T_N}\right)^{\alpha}\right]^{\beta}\left[1+\frac{C}{T-\theta}\right] \qquad (2)$$

Shown as the solid line in Fig 6, the extracted values $f_0$ = *14.1(2),* α = 4.3(6), and β = 0.23(5) are in reasonable agreement with those determined from the simple mean-field fit for Y-227, with $T_N$ = 122(2) K. The last factor in Eq 2 describes contribution of the polarized Curie-like $Yb^{3+}$ moments to the total magnetic field at the muon stopping site, with phenomenological parameters found to be $C$ = 0.8(5) K and $\theta$ = -7.5(2) K.

We have also studied the ZF and LF depolarization on the EMU spectrometer at ISIS to exploit its sensitivity to slow muon depolarization processes. Oscillations observed on GPS lie above the bandwidth of the EMU detector and thus only the slow-relaxing components are observable. The resulting asymmetry curves can be described by a simple exponential function,

$$A(t) = A_S exp(-\lambda t) \qquad (3)$$

where $A_S$ is the initial asymmetry and left as a fit parameter to account for the apparent loss of asymmetry due to the onset of oscillations. After correcting for background contributions and normalizing to full asymmetry, the low temperature value of $A_S$ was found to be nearly identical to η as determined from data taken on GPS. Finally, as a test of our model proposed in Eq. (1), the longitudinal field dependence of $A_S$ was measured for both samples at 1.8 K, as shown in Figure 7. For polycrystalline samples, the dynamic or longitudinal fraction of the total asymmetry (η or $A_S$) should vary as a function of applied longitudinal field as:

$$A = \frac{3}{4} - \frac{1}{4b^2} + \frac{(b^2-1)^2}{8b^3}\ln\left|\frac{b+1}{b-1}\right| \qquad (4)$$

with $b = \langle \boldsymbol{B_{loc}} \rangle / B_{ext}$, the ratio of the internal field to the applied external field [18]. Performing a fit of the data to Eq. 4, shown as the solid line in Fig 7 for Y-227, we find $\langle \boldsymbol{B_{loc}} \rangle$ = 950(200) G for Y-227 and 1130(150) G for Yb-227. These results are consistent with the values extracted from the low temperature muon spin precession frequencies, and the good fit of the model to our data confirm that nearly 100% of the sample has uniform long-range ordering.

The temperature dependence of λ is shown in the lower panel of Fig 5 for Y-227 for both spectrometers. The rates are somewhat higher for data taken on EMU than GPS, and could either reflect the difficulty of measuring such small depolarization rates at continuous muon sources such as PSI and/or a very slight underestimation of the background signal in the ISIS data. However, we note these rates are $10^3$ – $10^4$ times smaller than Λ or ω, therefore such disagreement between spectrometers produces negligible error in describing the temperature dependence of the oscillatory component. A small increase in λ is observed beginning at 190 K in Y-227 and correlates with bifurcation of the magnetization. This is similar to the behavior observed near the transition to the disordered magnetic state at $T$ = 120 K in Nd-227 [8]. As λ represents depolarization primarily from dynamic mechanisms, this increase signifies a modification of the local fluctuation spectra which may arise from any number of mechanisms in this frustrated system, including a growing presence of short-range order as suggested to occur in other iridate compounds [19].

Although λ exhibits a peak near the onset of spontaneous oscillations, there is no sign of paramagnetic divergence as typical for slowing down of spin-lattice fluctuations near a transition to a magnetically ordered or glassy state. Below 150 K, λ shows only weak temperature dependence and is clearly non-zero at 1.5 K indicating small contributions from either quantum fluctuations or a singular contribution of the excitation spectra near zero energy as proposed for Eu-227 [5] and $Y_2Mo_2O_7$ [20, 21].

The qualitative behavior of λ is remarkably similar between Y-227 and Yb-227 near the onset of oscillations, indicating that the local magnetic field is not substantially altered by the presence of additional moments. For Yb-227, below $T^*$ we observe a rather dramatic decrease of λ indicating substantial slowing down of fluctuations and coinciding with the increase of the static local magnetic field. As a probe of this behavior, the longitudinal field dependence of λ was measured at 1.8 K in an applied magnetic field up to 3 kG. Shown in the inset of Fig 8, λ is well described by a modified Redfield formula which describes depolarization due to dynamic fluctuations with a single relaxation channel with characteristic correlation time $\tau_c \gg (\gamma_\mu\Delta)^{-1}$ [14]

$$\lambda = \frac{\gamma^2\Delta^2\tau_c}{1+\omega\tau_c} + \lambda_0 \qquad (5)$$

where $\omega/\gamma_\mu = B_{ext}$, and Δ is the width of the field distribution of the fluctuating moments at the muon site. The constant $\lambda_0 = 0.05(2)$ MHz is used here as a correction for other field independent relaxation channels. Fitting the data with Eq. 5 yields Δ = 20(2) G and $\tau_c = 1.0(5) \times 10^{-9}$ s; this value of Δ indicates small variations of the local field relative to $\langle \boldsymbol{B}_{loc} \rangle$. Our extracted value of $\tau_C \sim 1$ ns is several orders of magnitude larger than typical paramagnetic systems, and is representative of systems with an energy barrier to fluctuations, such as a moment in a strong polarizing field or as observed for example in single molecule magnets [22].

## IV. Discussion

We have examined two species of the iridate pyrochlore family via magnetization, elastic neutron scattering and μSR measurements. At low temperatures both compounds clearly show similar static local fields at the muon-stopping site attributed to commensurate long-range order. The lack of long-range order from neutron scattering measurements in Y-227 and Nd-227 combined with recent studies of other frustrated Ir-based compounds [23] suggests that the large absorption cross section of the Ir-nuclei and inherently small ordered moments prevent the observation of any spin order in polycrystalline samples. Future neutron studies on single crystals may be necessary then to accurately determine the ground state configuration of the $Ir^{4+}$ moments.

We now turn to the low temperature results. Assuming that the muon-stopping site is the same for all the insulating iridate compounds examined (Y, Yb, Eu), our results suggest that the magnetic structure is the same for all three at low temperatures. As neutron scattering is not able to provide a reliable determination the magnetic structure, we have attempted to obtain some insight into this by modeling the internal field of Y-227 based on simple dipolar coupling of the muon to the $Ir^{4+}$ moments. Possible muon-stopping sites have been deduced by Dunsiger for the isostructural $Y_2Mo_2O_7$ using Ewald's method to solve the electrostatic potential of the crystal [43], and by McClarty *et al.* [25] from experimentally measured muon rotation frequencies in insulating $Gd_2Sn_2O_7$ where the Palmers-Chaulker configuration was assumed for the ground state. The muon-stopping sites, normalized to the appropriate lattice parameter, were found to be [0.16, 0.16, -0.17] (site 1) and [0.4402, 0.5005, 0.5625] (site 2) for $Y_2Mo_2O_7$ and $Gd_2Sn_2O_7$, respectively. We have calculated the magnetic field at these locations based on dipolar fields from the $Ir^{4+}$ moments, neglecting any exchange interaction [26], using a cube of five unit cells per side and lattice parameter experimentally measured in this work. The $Ir^{4+}$moment was chosen to be 1 $\mu_B$ which is higher than the estimated moment based on our neutron diffraction results; however it represents the largest value for the $J$ = 1/2 $Ir^{4+}$ and will therefore serve as a upper bound for the following results. The resultant field has been calculated for the possible lowest energy states proposed in Ref 2. These include the non-collinear all-in/all-out and 2-in/2-out configurations, two collinear ferromagnetic configurations aligned along either the (111) or the (100) directions (FM-(111) and FM-(100), respectively), and one co-planar state with orthogonal spins due to the indirect Dzyaloshinsky-Moriya interaction, or IDM. We report the magnitude of the field for each configuration in Table 1 with an error due to imprecision of the stopping site as determined by the maximum variation of the field over a spherical of radius of 10% the interatomic distance between the oxygen and iridium ions nearest the proposed stopping sites.

For both sites we find that the non-collinear and FM-(111) configurations overestimate the size of the magnetic field relative to that measured by ZF-μSR, while the co-planar IDM and FM-(100) configuration underestimate this value by a significant amount. As these values represent the upper bound of the local magnetic field we can rule out the IDM and FM-(100) as the correct ground state configurations. However, we are unable at this time to distinguish between non-collinear sites and FM-(111) configurations based on the imprecision in the determined muon stopping-site. Future studies of single crystal samples would allow for better estimates of the stopping site symmetry and thus allow for a better determination of the ground state structure.

We next turn to the unusual behavior of Y-227; the onset of strong magnetism near 190 K in bulk susceptibility together with lack of spontaneous oscillations in ZF-μSR indicates that there is no long-range order at this transition. This is also not a transition to a traditional spin-glass state as in the isomorphic pyrochlore $Y_2Mo_2O_7$ [21,22], since there is no divergence in the longitudinal relaxation rate λ characteristic of a critical slowing down of fluctuating moments at temperatures just above the freezing temperature. Based on our observations, we propose that this is the same phenomenon as observed in Nd-227 [8], in which a magnetic state lacking long-range order was shown to exist over a wide temperature range, with long-range order as determined from ZF-μSR not setting in until a much lower temperature than the bifurcation temperature $T_M$. Previous measurements of Nd-227 [8, 13] suggest that the Nd-Ir exchange interaction is far too weak to induce this type of disorder, and that it should rather be due to intrinsic interactions of the Ir-sublattice alone. This idea is confirmed by the presence of a such a state with nonmagnetic $Y^{3+}$.

This existence of unusual magnetically disordered phase near $T_M$ is in contrast to the behavior observed for Yb-227, which exhibits a comparatively simple transition from paramagnetic behavior to long-range order with no intermediate phase. Taking these results with the previously reported results for Nd-227 [8], we conclude that the formation of the frustrated or short-ranged ordered state and its stability against transition to long-range order is closely associated with *A*-site radii $R_A$, as $R_{Yb} < R_Y < R_{Nd}$. It has been suggested separately that variations of the lattice parameter affect not only the conduction bandwidth, but also correlation and Ir-Ir exchange energy [1-3, 27,28], both of which in turn determine transport properties and ground state configuration. We propose that competition between these parameters as determined by the *A*-site radii lead to the glassy/short-range ordered phase for a finite range of relative energy scales. It would be of great interest therefore to continue these comprehensive measurements for the remainder of the rare-earth series to verify this prediction and gain further insight into this mechanism.

## V. Conclusion

In summary, we observe long-range order in both Y-227 and Yb-227 which is likely the same for all insulating phases Furthermore, we report an unusual disordered precursor spin phase in Y-227 which suggests that bifurcation, metal-insulator, and long-range ordering need not occur at the same temperature as previously reported. Our results support the predictions that correlation strength can be controlled by choice of *A*-site element and has direct consequences on the magnetic structure, including the stabilization of an intermediate glassy or short-range ordered phase in Y-227 as observed in Nd-227. While these observations have provided significant insight into the underlying complexity of these materials, fabrication of large volume, high-quality single crystals is still necessary for the conclusive determination of the magnetic structure. Comprehensive studies of the remaining members of the iridate family are vital to answering the remaining questions and to assist with the development of future theoretical models.

M.J.G. and S.D.W. would like to acknowledge very helpful discussions with Ying Ran. This work was supported in part by National Science Foundation Materials World Network grant DMR-0710525 (M.J.G.) and by NSF CAREER award DMR-1056625 (S.D.W.). Muon experiments were performed at the ISIS Muon Facility at the Rutherford Appleton Laboratories (UK) and the Swiss Muon Source at the Paul Scherrer Institute (Switzerland). Part of this work was performed at Oak Ridge National Laboratory High Flux Isotope Reactor, sponsored by the Scientific User Facilities Division, Office of Basic Energy Sciences, U.S. Department of Energy.

TABLE 1. The resultant field at two possible muon-stopping sites for various configurations of the $Ir^{4+}$ sublattice as described in the text.

| | all-in/all-out | 2-in/2-out | FM-(111) | FM-(100) | IDM |
|---|---|---|---|---|---|
| Site 1 | 1600 (200) G | 1500 (200) G | 1400 (100) G | 700 (100) G | 300 (100) G |
| Site 2 | 1250 (100) G | 1250 (100) G | 1200 (100) G | 500 (50) G | 620 (50) G |

**List of Figures**

Figure 1. XRD patterns for (a) Y-227 and (b) Yb-227 with Rietvald refinement. (c) Diffraction pattern from elastic neutron scattering of Y-227 at 3 K.

Figure 2. Temperature dependent susceptibility of (a) Y-227 and (b) Yb-227 with inset showing enlargement of data around bifurcation of Yb-227 near 130 K. Temperature dependent resistivity of (c) Y-227 and (d) Yb-227; insets show data on log-log scale.

Figure 3. (a) Representative short time asymmetry data taken from GPS for Y-227 with curves offset for clarity and solid lines representing fits to Equation (1). (b) FFT transforms of the asymmetry data smoothed to highlight evolution of the peak

Figure 4. (a) Representative short time asymmetry data taken from GPS for Yb-227 with curves offset for clarity, and solid line representing fits to Equation (1). (b) FFT transforms of asymmetry data.

Figure 5. Temperature dependence of the parameters extracted from fitting Eq (1) for Y-227 TOP: $\omega_\mu/2\pi$ with $\eta$ inset; the solid line is a mean field fit, as described in the text. MIDDLE: $\Lambda$, solid lines are guides for the eyes. BOTTOM: $\lambda$ from both GPS and EMU

Figure 6. Temperature dependence of parameters extracted from fitting Eq (1) for Yb-227. TOP: $\omega_\mu/2\pi$ with $\eta$ inset; the solid line is afit to Eq. (2), as described in the text. MIDDLE: $\Lambda$ solid lines are guides for the eyes. BOTTOM: $\lambda$ from both GPS and EMU

Figure 7. Field dependence of the initial asymmetry taken on the EMU spectrometer at 1.8 K for Y-227 and Yb-227. The solid line is the fit of Eq (4) to the Y-227 data.

Figure 8. Field dependence of $\lambda$ for $Yb_2Ir_2O_7$ taken from the EMU spectrometer at 1.8 K. The solid line represents the fit of Eq (5).

Figure 1

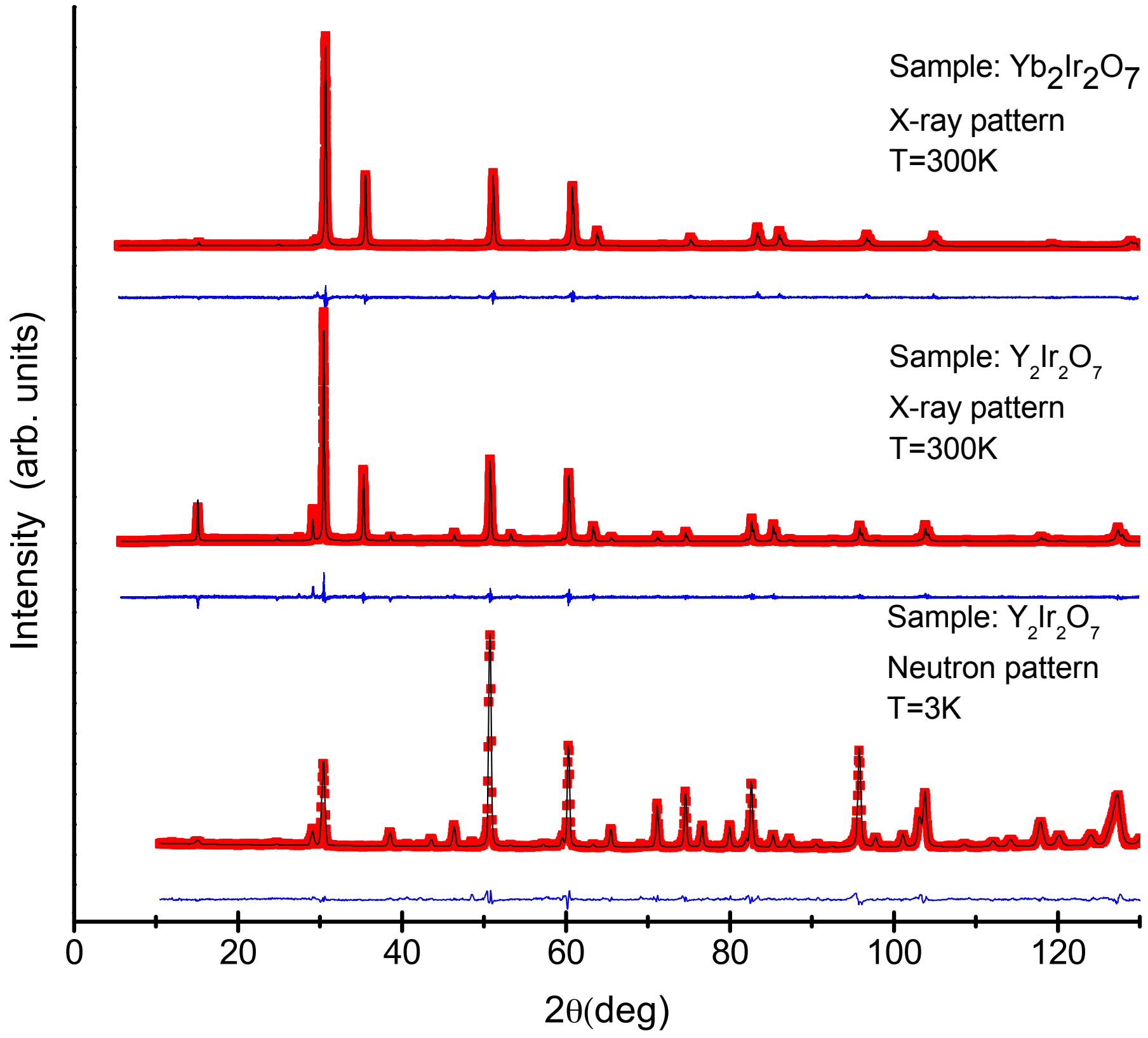

Figure 2

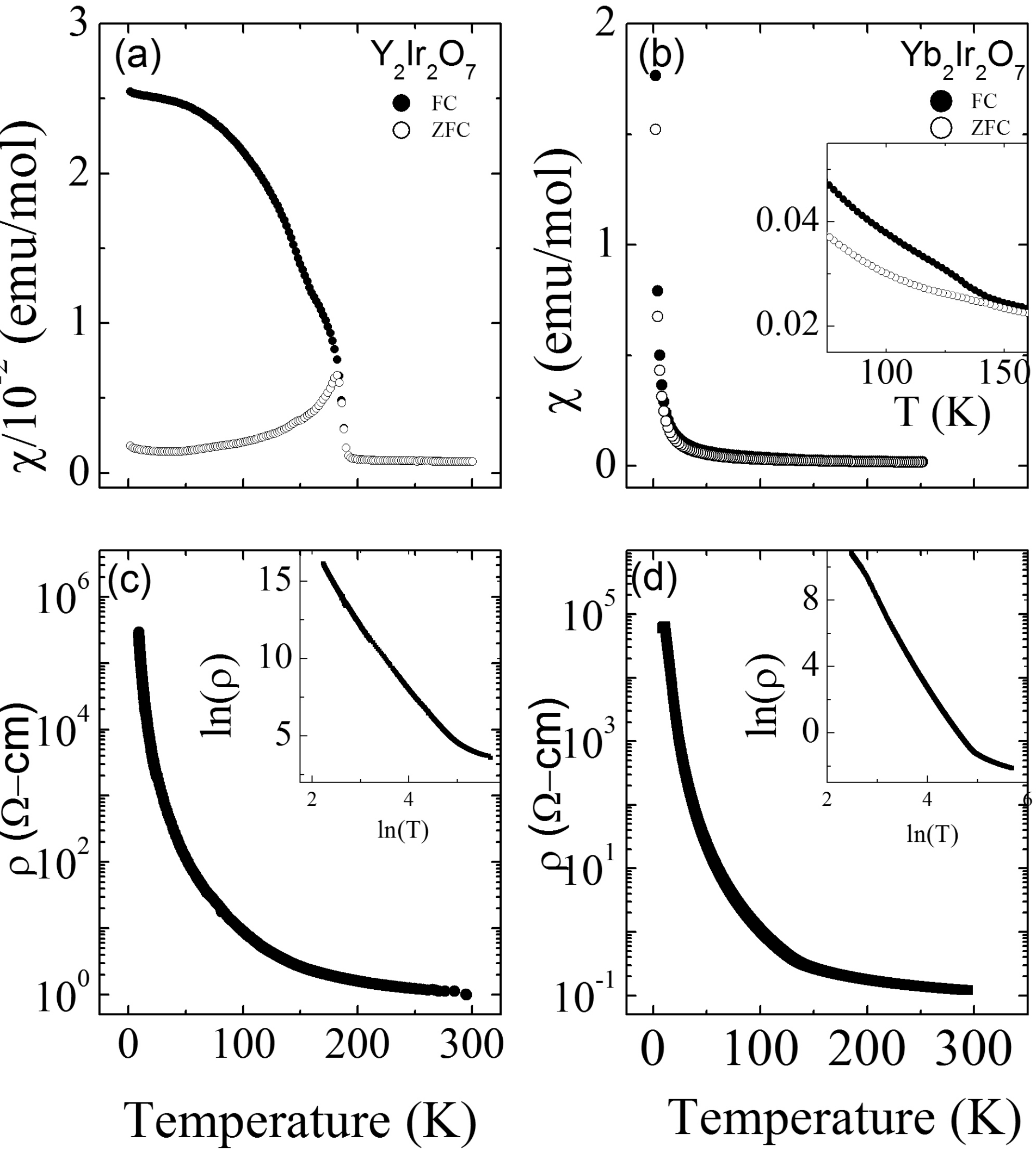

Figure 3

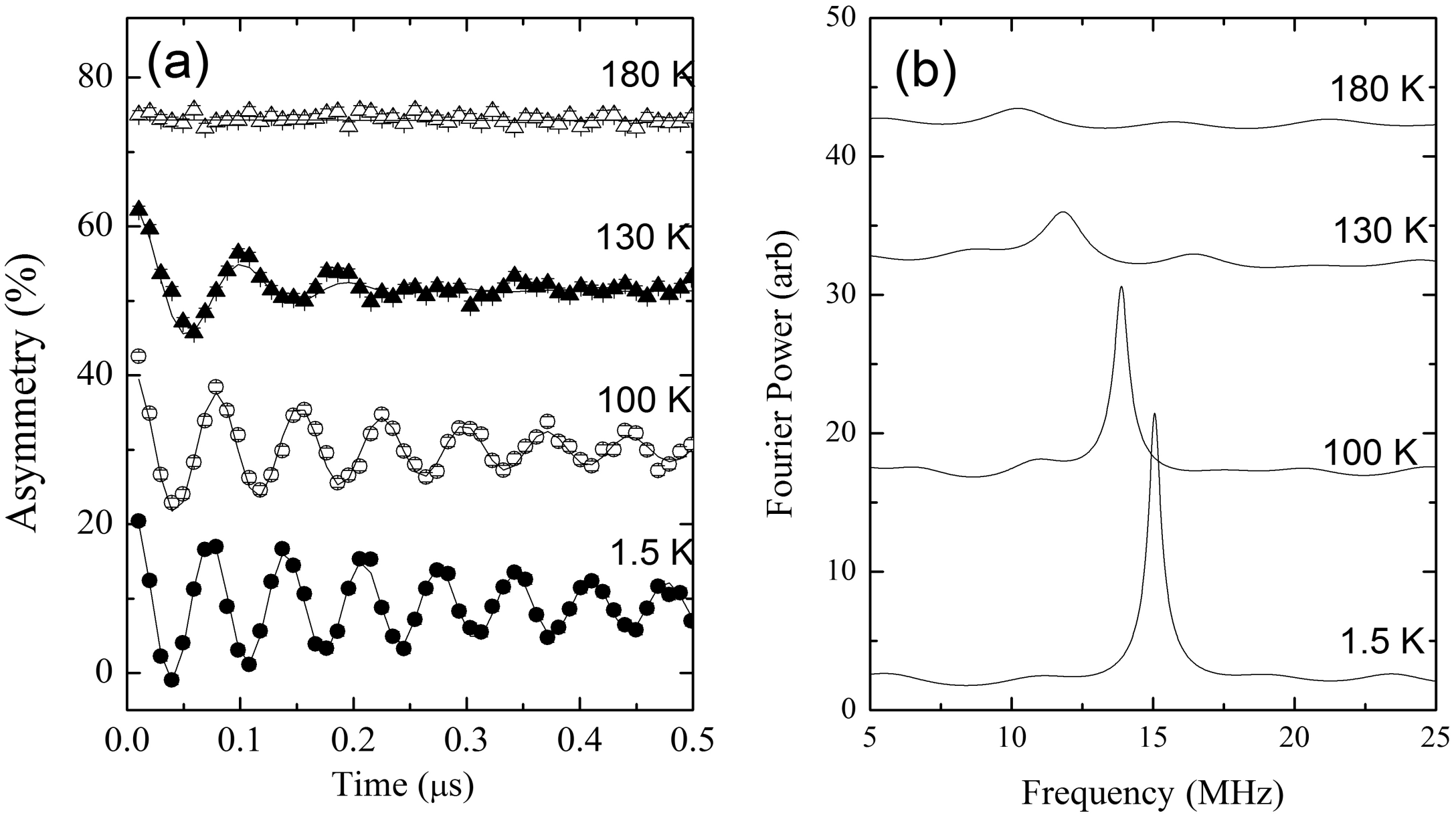

Figure 4

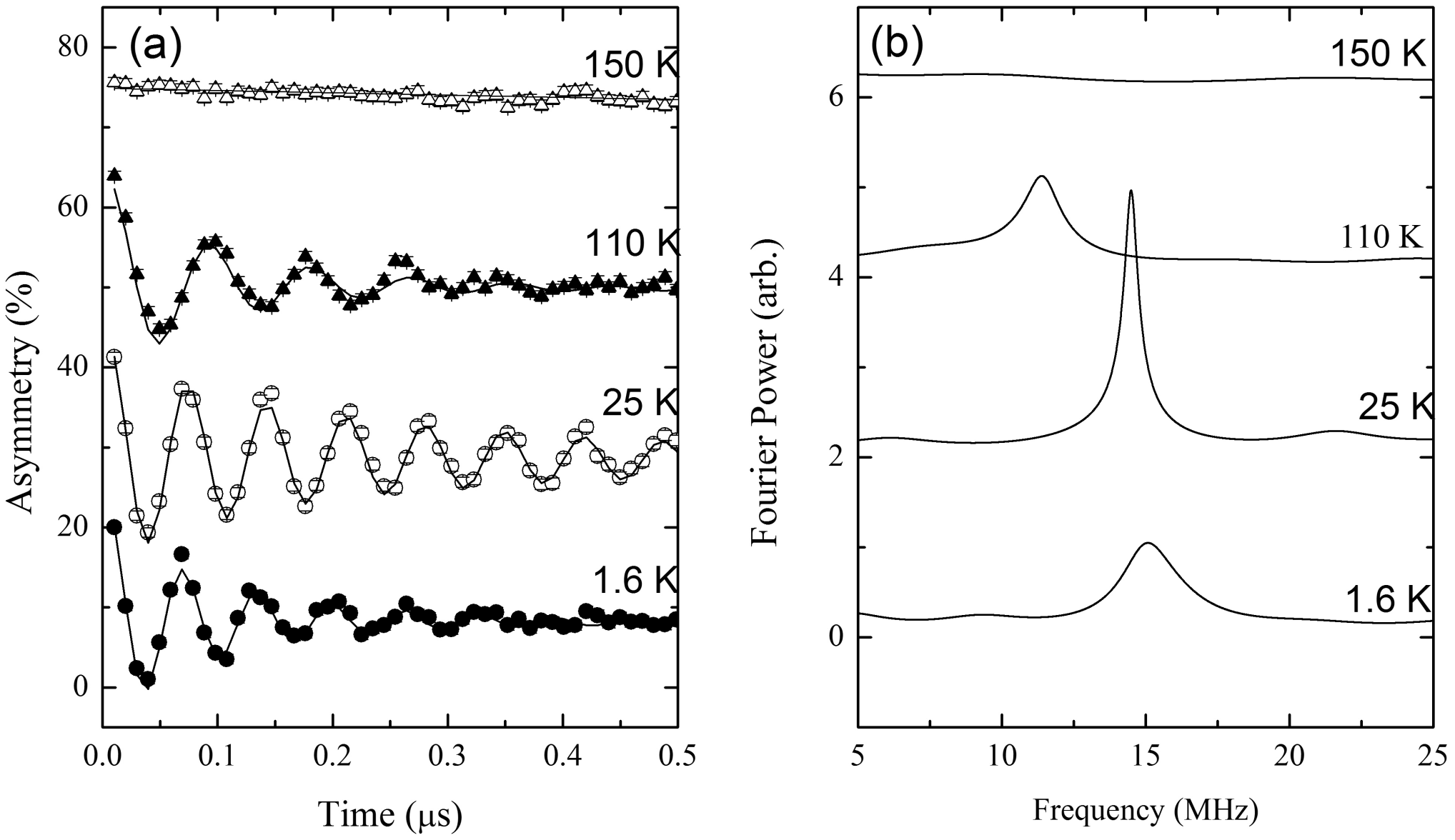

Figure 5

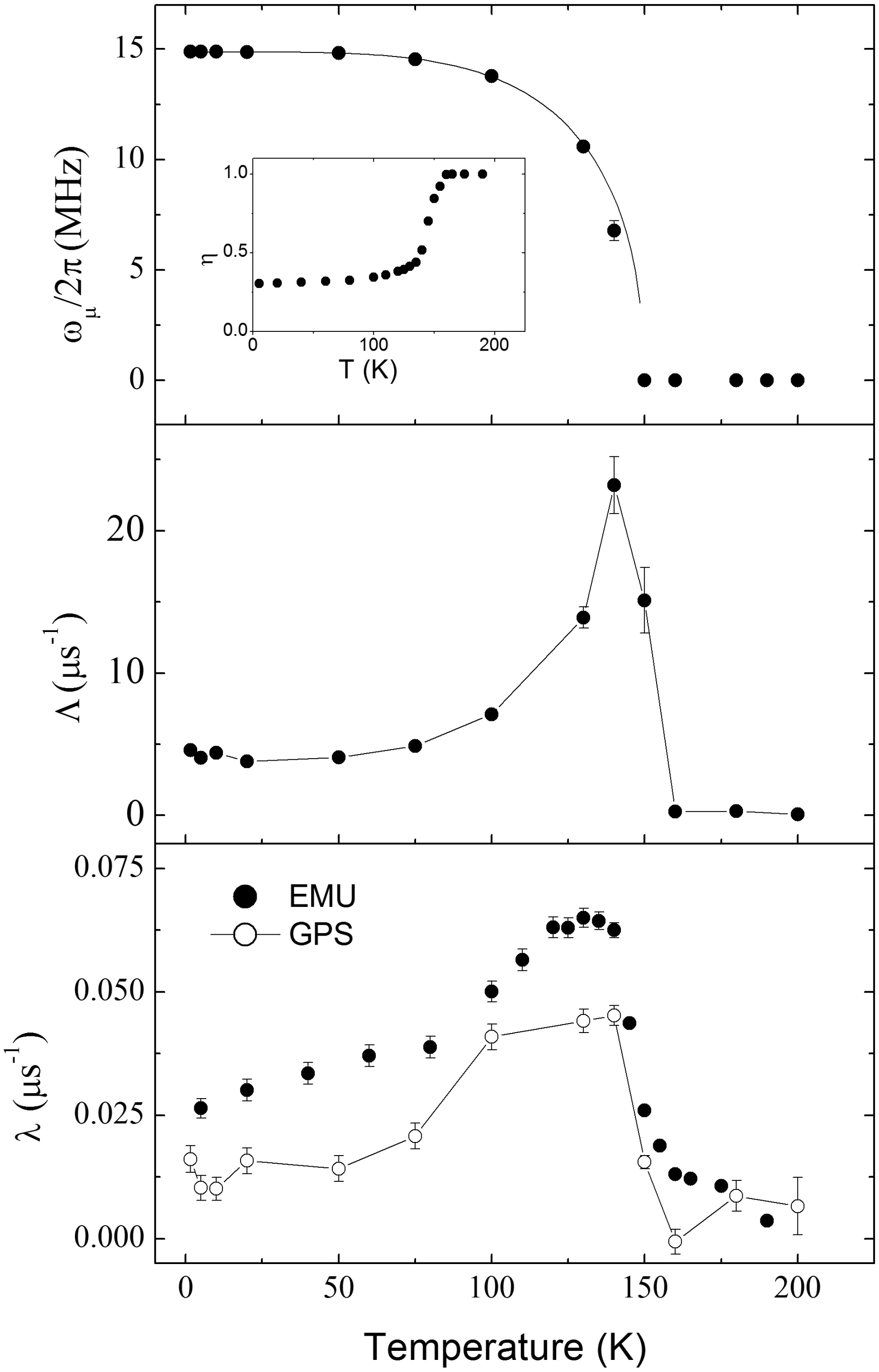

Figure 6

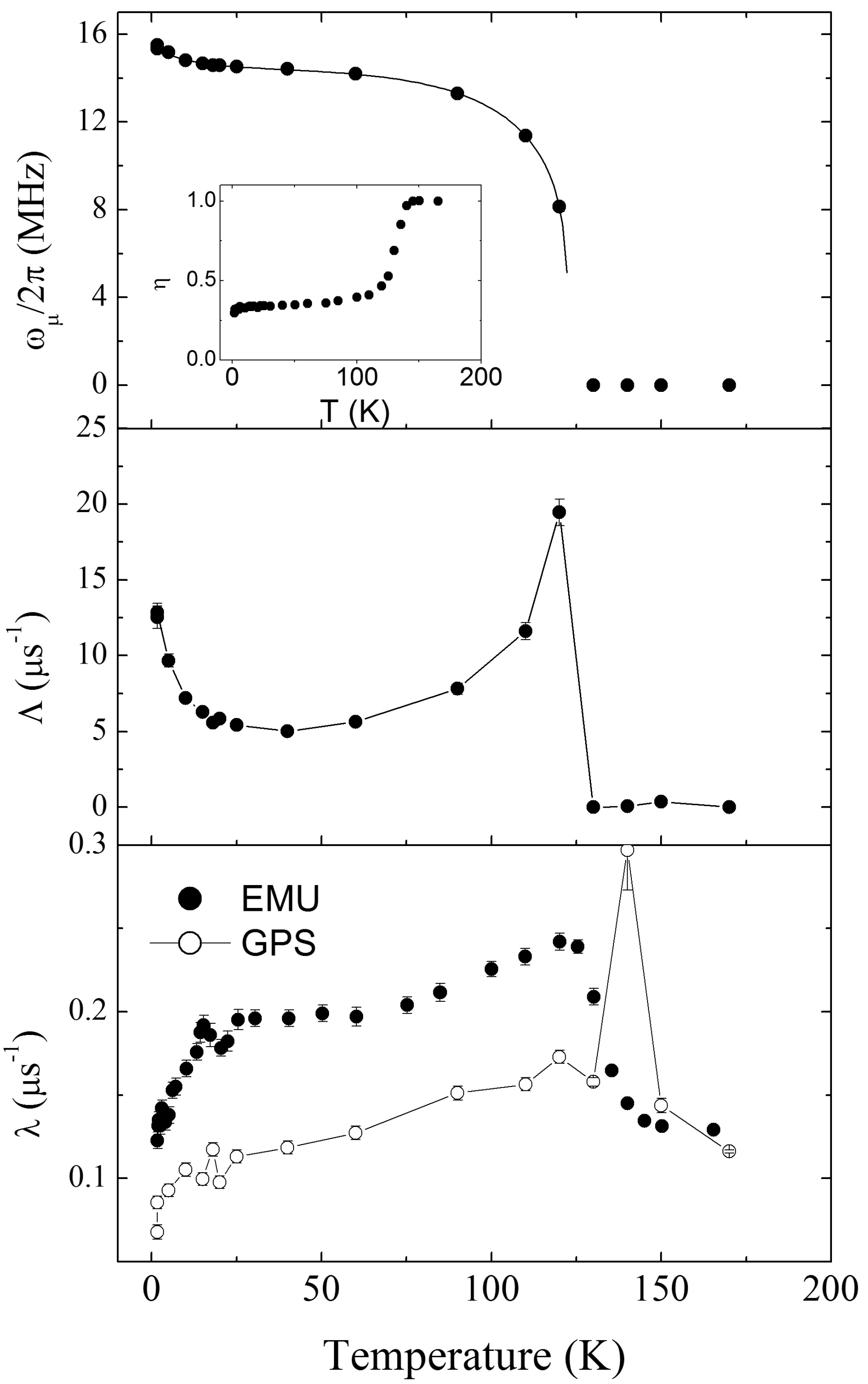

Figure 7

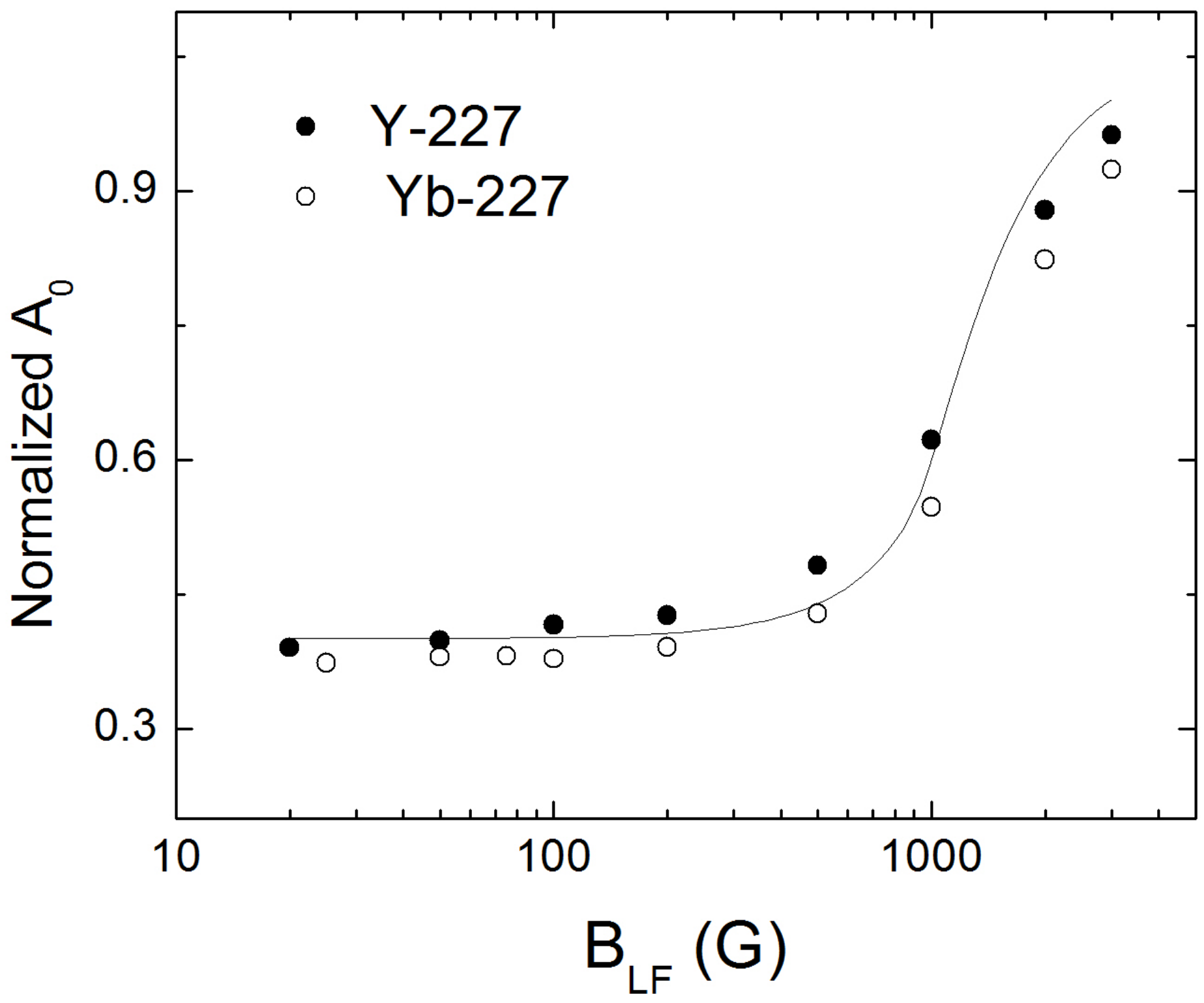

Figure 8

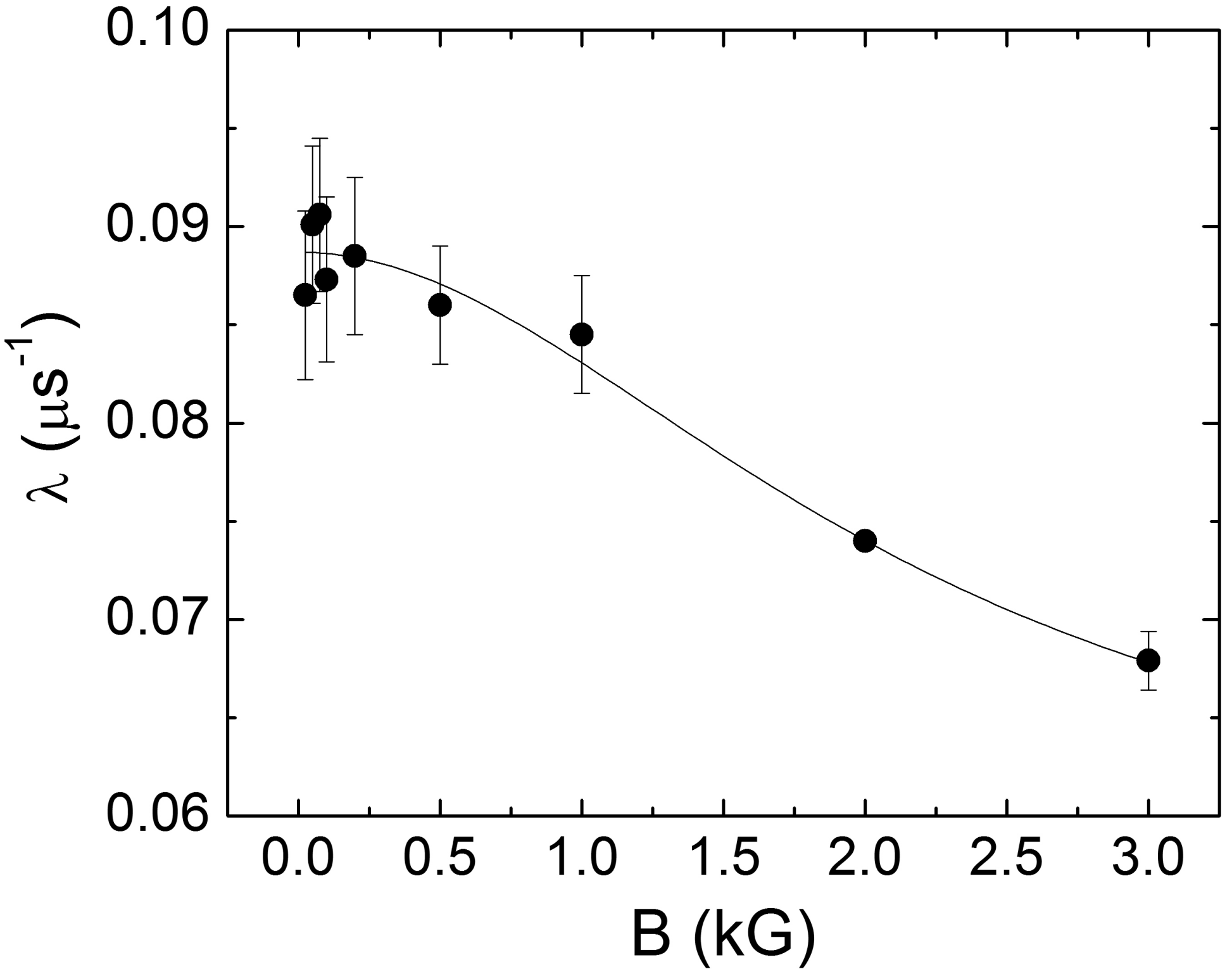